# Gender-Wise Perception of Students Towards Blended Learning in Higher Education: Pakistan

Saira Soomro [1], Arjumand Bano Soomro [2,4], Tarique Bhatti [3] and Yonis Gulzar [4]

[1] Department of Distance Continuing and Computer Education, Faculty of Education, University of Sindh, Hyderabad, Pakistan
[2] Department of Software Engineering, Faculty of Engineering and Technology, University of Sindh, Jamshoro, Sindh, Pakistan
[3] Department of Psychological Testing Guidance & Research, Faculty of Education University of Sindh, Hyderabad, Pakistan
[4] Department of Management Information Systems, College of Business Administration, King Faisal University, Al Ahsa, Saudi Arabia

## ABSTRACT

Blended learning (BL) is a recent trend among many options that can best fit learners' needs, regardless of time and place. This study aimed to discover students' perceptions of BL and the challenges faced by them while using technology. This quantitative study used data gathered from 300 students enrolled in four public universities in the Sindh province of Pakistan. The findings show that students were comfortable with the use of technology, and it has a positive effect on their academic experience. The study also showed that the use of technology encourages peer collaboration. The challenges found include: neither teacher support nor a training programme was provided to the students for the courses which needed to shift from a traditional face to face paradigm to a blended format, a lack of specialised skills in laboratory assistants for the courses with a blended format and a shortage of hi-tech computer laboratories/computer units to run these courses. Therefore, it is recommended that authorities must develop and incorporate a comprehensive mechanism for the effective implementation of BL in the teaching-learning process. Heads of departments should also provide additional computing infrastructure to their departments.

## 1. Introduction

Higher education institutions are important in the development of skilled and highly qualified people who can play an active role in all sectors of the national economy. According to a World Bank report from 1994, 'higher education institutes are responsible for providing the knowledge and skills required for a person to achieve a target position in different sectors of professions, such as government and business. In most countries, higher education institutions play a vital role in all sectors of the national economy and social sectors as well'. In this contemporary era of a 'new normal' world that is technology-dependent, students will only succeed if digital technology is properly used by teachers in higher education (Okaz, 2015).

Universities are the backbone of a nation's economic growth as they produce new youthful members of the workforce with new digital literacy. Higher education institutions are places where students of different colours, genders, races and cultures meet. Higher education teachers with professional expertise in blended earning have an impact on students' learning (Impedovo and Khatoon Malik, 2019). Economic growth and prosperity are only possible when universities adapt and bring new developments to the fields of science and technology. With the technology, new avenues of communication in the digital world have been opened. To impart this digital literacy to students, higher education institutes need to accept revolutionary changes in the field of digital technology. At this point, higher education institutions are in a challenging position to deal with these revolutionary digital changes. For this, the higher education system must be dynamic to accept the changes of the current digital world, and new students have to cope with the challenges of the new digital world. With the new technological revolution, drastic changes have impacted the overall educational system, particularly higher education in developed countries. The concept of BL with advance knowledge of digital technologies has flourished in educational institutions in the developed world. The reduction in lecture-based teaching methods and dependence on printed materials has made the use of BL (the combination of technology and face to face traditional classroom approach) possible. Innovation, flexibility, activity and collaboration is made possible due to the BL approach in teaching-learning. With the facilitation of BL, students can access online platforms and pre-recorded lectures anytime and anywhere. The BL approach is also cost-efficient in higher education and has brought a new dimension to peer-to-peer interaction and peer-to-faculty interaction (Garrison and Vaughan, 2008).

A developing country such as Pakistan has a low budget for the education sector. Despite this challenge, the Higher Education Commission and Federal Ministry of Education have tried to utilise funds for the improvement of education standards. Fortunately, free MOOCs resources are available. In Pakistan, BL has brought a variety of teaching and learning possibilities combined with face to face teaching. BL provides a user-friendly open learning management system with an option of interaction and communication between peers. Students are now free to work online and check the uploaded lectures and video tutorials from their teachers. Teachers can also conduct assessments with online assignments and quizzes (Memon and Rathore, 2018). In Pakistan in 1974, the Allama Iqbal Open University (AIOU) was established to provide non-formal education via the correspondence model of distance education. AIOU has now applied and tested various methods of media and technology. After AIOU, the Virtual University (VU) of Pakistan, established in 2000, created another technology-based initiative. The scope of technological use by these two universities is still limited to distance-learning students. The initial framework of the Open Learning Institute of Virtual Education (OLIVE) as a concept of e-learning was first coined by AIOU in Pakistan back in 2000. The VU now has four on-air channels on the Pakistan communication satellite network.

Besides these two, no other public and professional universities in Pakistan have adopted the advanced technological system fully to include BL. As per a report from the Higher Education Commission (HEC) of Pakistan in 2000, the factors responsible for not adopting technological tools are barriers of language, current state of use of ICT penetration, computer literacy and hesitation to move away from traditional learning methods. With the establishment of the HEC of Pakistan in 2000, many steps to upgrade the standard of higher education have been taken. The HEC of Pakistan has taken various initiatives to support and improve higher educational institutions, including online lecturing and net-meeting using an IP-based video conferencing system, broadband facility, national digital library and the Pakistan Education and Research Network.

The HEC of Pakistan has increased ICT use in the teaching and learning process of its faculty and students by incorporating various training programmes. It has been working continuously on various projects to provide adequate ICT infrastructure and content development for students. It has also established the Directorate of Distance Education, which encourages e-learning and distance learning. A study was conducted on teachers' attitudes towards BL in Pakistan (Soomro, Soomro, Bhatti and Ali, 2018); the findings of this study show that there is the need for a study to be conducted to explore the thoughts of higher education students.

Through the adoption of a BL model in the university system, Pakistan will have many advantages; for example, students can use online platforms, pre-recorded lectures, peer-to-peer and peer-to-faculty interaction instructional models and real-time discussions with experts and experienced people online.

The remainder of the paper is organised as follows: section two contains the rationale of the study. The objectives of the study are set out in section three and hypotheses are mentioned in section four. In section five, the previous studies relating to research domain are reported and discussed. In section six, the methodology of this research is explained. Section seven contains the analysis and interpretation of data. Discussion and conclusion are mentioned in section eight, and section nine suggests recommendations from the research.

## 2. Rationale

Education paradigms have changed with the rise of blended learning (BL), and the use of BL has had far-reaching impacts on higher education. BL promotes individual learning at the student's own pace, peer collaboration and communication. It introduced a type of learning, where traditional class mode can be kept along with application of ICTs. BL gives face to face lessons along with virtual classrooms. In this context, the research topic under investigation is 'Use of BL at university level as perceived by the students'.

## 3. Objectives

- To find out the opinions of students about the use of technology in BL.
- To investigate the obstacles faced by students at general public universities in the use of BL.

## 4. Hypotheses

- $H_0$: There is no significant difference in the perceptions of male and female students about the use of blended learning for learning purposes.
- $H_1$: There is a significant difference in the perceptions of male and female students about the use of blended learning for learning purpose.

## 5. Literature Review

E-learning is a fully online mode; BL is another method that combines traditional face to face and online modes with technology-based tools that teachers and students are given to create an ideal environment for effective learning. Students' feedback and results are a true assessment of BL as it covers a wide range of students. Teaching through BL takes less time than the traditional setup. BL provides both modes of teaching, and it becomes cost-effective when technology and instruction are integrated and cater for different students' learning styles. BL is the combination of instructional strategies, teaching pedagogies and theories of learning (McGrath, 2013).

The concept of BL has emerged from the existence of E-learning or fully online instruction. In the 21st century, the use of technology has become mandatory in all disciplines of higher education. As Littlejohn and Pegler (2007) discussed, traditional methods of teaching do not fit every student's needs, and with web-based education higher education becomes the platform for students to meet different challenges. Studies carried out by Bonk and Graham (2012) and Mohandes, Dawoud, Al Amoudi and Hussain (2006) checked the satisfaction levels of learners in online learning systems. The results showed that most of the participants favoured BL. Another study conducted by McGrath (2013) showed that BL offers effective learning, access to technology and a reduction in cost. However, So and Brush (2008) discussed the dissatisfaction felt by teachers and students because it does not meet the needs of a fully online learning system.

King and Cerrone Arnold (2012) conducted a study on success factors for BL. She found motivation, communication design and course design the most important factors to be considered for BL success. Kistow (2011) found that the majority of students favoured the BL mode because of its accessibility, flexibility and suitability. BL allows students a degree of freedom to interact with their peers. According to Graham et al., (2005) and Osguthorpe and Russell (2003), BL provides flexibility and accessibility for higher education. According to Qureshi, Ilyas, Yasmin and Whitty (2012), a country such as Pakistan will not benefit from BL until, and unless, there is an electric power failure and English language barriers are persistently present in higher education institutes.

Singh (2003) provided different learning styles relevant to the nature of their subjects for different students' learning needs. According to him, with the use of different teaching strategies and tactics, students can choose the method of learning that suits them best. According to Collis, Betty and Huib Bruijstens (2003) and Morgan (2002), BL provides students with 21st century skills, i.e. peer collaboration and problem-solving skills, due to its pedagogical structure. BL gives the best results by combining different strategies of traditional face to face methods and distributed systems.

A mixed-method study was conducted on levels of BL implementation in the province of Sindh, Pakistan. From the analysed data and interviews, it was concluded that sample universities secured the initial stage of a framework given by Graham et al., (2013). The participants cited various reasons for not being able to implement BL properly, such as limited computing infrastructure in comparison to student population, no training or official guidance for implementation of BL in classrooms, no model of BL nor any international established framework adopted and no course being designed to offer students the benefits of BL (Soomro et al., 2018).

Another study conducted in Pakistan and East Africa at the same time cited some reasons for discouraging students to continue with BL. Most of them stated that electricity shutdown was the major challenge for them continuing with this method of learning. Some of them felt that this form of learning would cause them to miss out on interaction in a face to face classroom community and they may lose their motivation. The cultural and social aspect of learning plays a major role in the successful implementation of BL (Rizvi, Gulzar, Nicholas and Nkoroi, 2017).

## 6. Methodology

The research study has followed a quantitative approach based on 128the survey. The survey questionnaire was developed and administered in the sample population. There are a number of public sector universities in the Sindh province, out of which four general public sector universities were selected, namely University of Karachi, University of Sindh Jamshoro, Shaheed Benazir Bhutto University Benazirabad and Shah Abdul Latif University Khairpur.

There were two reasons for selecting these universities. First, these are general public sector universities of the Sindh Province, in which initiatives of the BL mode of learning have been introduced. Second, the majority of students at these universities are studying various fields of science and the arts.

There were 55964 students enrolled at these four general public sector universities. For sampling, four technologically rich departments, two from the Social Sciences Faculty and two from the Natural Sciences Faculty, were selected through purposive sampling, and 5% of students (male and female) from each of these departments were randomly selected. The total number of students selected for the study was 332 of either gender. The questionnaire, based on the 5-point Likert scale, was used to collect data from the students. Simple percentages and mean scores were used for data analysis. The null hypothesis was tested through t-test for alpha at 0.05 significance level.

For determining the reliability of the questionnaire, the Cronbach Coefficient alpha technique was used. The reliability came out as 0.812. For the validation of the questionnaire, language, content and structure were reviewed and checked by the experts. Simple percentages and mean scores were used for simple descriptive data analysis. The inferential statistics were used to test the null hypothesis. The T-test was used to test the hypothesis and to see the mean difference among two groups of male and female students, as suggested by Pallant (2010).

## 7. Data Analysis and Interpretation

The questionnaire was administered among 332 students and answered by 300. Out of 332 students, 179 were male and 153 were female. The response rate for male students was 84%, that is 152 responded out of 179. The response rate for female students was 96%, which means that 148 participants out of 153 answered the questionnaire.

### 7.1. Item-Wise Analysis Through Percentages and Simple Mean Scores:

The responses gathered in terms of strongly agree (SA) and agree (A) have been accumulatively reported for the following tables. Also, the responses for 'disagree' (DA) and 'strongly disagree' (SDA) are also collectively mentioned in terms of percentage. 'Undecided' (UD) have been kept separate from both positive and negative trends, to see the other side of perception from the sample.

Table I shows that the majority of students (60.67%) agreed that they have technology equipped classrooms and laboratories, while 34.67% of students disagreed and 4.67% remained undecided. The mean score is 3.3, which shows that the majority of students agreed that they have technology equipped classrooms and laboratories for their learning process. However, technology-equipped classrooms were mostly available in natural science faculties and less in social science faculties.

Table 1: availability of technology-equipped classrooms and labs

| Scale | SA | A | UD | DA | SDA | Mean |
|---|---|---|---|---|---|---|
| Freq. | 48 | 134 | 14 | 68 | 36 | 3.3 |
| %age | 16.00% | 44.67% | 4.67% | 22.67% | 12.00% | |

In table 2, 59.33% of the students were in favour of the statement that 'technical assistance is given by the teacher/support staff when it is needed', whereas 24.00% of students disagreed and 16.67% remained undecided. The mean score was 3.47. Therefore, it was concluded that, in the majority of departments, technical help is provided to students by the faculty and technical staff for the proper use of technology in the teaching-learning process. The large number of students made it difficult to reach, help and guide everyone.

Table 2: level of assistance given by the teacher/support staff

| Scale | SA | A | UD | DA | SDA | Mean |
|---|---|---|---|---|---|---|
| Freq. | 54 | 124 | 50 | 54 | 18 | 3.47 |
| %age | 18.00% | 41.33% | 16.67% | 18.00% | 6.00% | |

Table 3 shows that most of the students (80.66%) agreed that the use of technology affects their academic achievement positively, few students (9.67%) students disagreed and 10.00% remained undecided. The mean score was 4.04, and it is indicative that the use of technology is very beneficial for students, especially those from urban areas, as they can have access to advanced sources, books and other literature for their studies in a short space of time. However, students from remote areas face a lot of difficulties in the use of technological sources, and the majority of students at these general public sector universities are from remote areas.

Table 3: effects of using technology on student academic achievement

| Scale | SA | A | UD | DA | SDA | Mean |
|---|---|---|---|---|---|---|
| Freq. | 100 | 142 | 30 | 26 | 02 | 4.04 |
| %age | 33.33% | 47.33% | 10.00% | 8.67% | 0.67% | |

In table 4, it can be seen that a greater number of students (66.00%) agreed that the use of technology affects students' peer collaboration positively, whereas 13.34% students disagreed and 20.67% students remained undecided. The computed mean score was 3.69, which affirms that the use of technology is very useful and advantageous for students because they have access to peers and can easily share technology resources and e-books with other students. However, these types of collaboration need a high bandwidth to access such resources.

Table 4: effects of using technology on students' peer collaboration

| Scale | SA | A | UD | DA | SDA | Mean |
|---|---|---|---|---|---|---|
| Freq. | 56 | 142 | 62 | 32 | 08 | 3.69 |
| %age | 18.67% | 47.33% | 20.67% | 10.67% | 2.67% | |

The table 5 depicts that 48.34% of students agreed that their teachers use open education resources (OER), whereas 27.33% of students disagreed and 21.33% students remained undecided. The mean score remained 3.33, and it was concluded that the use of OER is advantageous for students as they can access shared e-books and enrol in Massive Open Online Courses (MOOCs).

Table 5: use of open education resources by the teacher

| Scale | SA | A | UD | DA | SDA | Mean |
|---|---|---|---|---|---|---|
| Freq. | 44 | 110 | 64 | 66 | 16 | 3.33 |
| %age | 14.67% | 33.67% | 21.33% | 22.00% | 5.33% | |

Table 6 shows that 44.00% of students affirmed that they use search engines to locate learning materials related to their subjects, and the same number of students disagreed, whereas 12.00% of students remained undecided. The computed mean score was 2.96. Therefore, it is concluded that students actively use technology for their subject relevant content.

Table 6: use of technology in searching for subject related contents

| Scale | SA | A | UD | DA | SDA | Mean |
|---|---|---|---|---|---|---|
| Freq. | 100 | 142 | 30 | 26 | 02 | 2.96 |
| %age | 33.33% | 47.33% | 10.00% | 8.67% | 0.67% | |

It can be seen in table 7 that the use of personal devices (e.g. cell phone, mp3 player, PDA) is easy due to their size and shape and provides e-resources to students that can affect their learning positively. This shows that most of the students (66.66%) responded that they used personal devices (e.g. cell phone, mp3 player, PDA) while learning, 16.66% students disagreed and 16.67% students remained undecided. The mean score was 3.68. A notable number of students also use these devices for personal use.

Table 7: use of personal devices (e.g. cell phone, mp3 player, pda) by students while learning

| Scale | SA | A | UD | DA | SDA | Mean |
|---|---|---|---|---|---|---|

| Scale | SA | A | UD | DA | SDA | Mean |
|---|---|---|---|---|---|---|
| Freq. | 70 | 130 | 50 | 34 | 16 | 3.68 |
| %age | 23.33% | 43.33% | 16.67% | 11.33% | 5.33% | |

Table 8 depicts that the majority of students (85.33%) agreed that they used social networking applications (e.g. Facebook, Twitter) which assist them during their learning process, very few (6.67%) students disagreed and 8.00% students remained undecided. The mean score was 4.18. The social networking applications support the e-groups and e-discussion forums that help students to collaborate and affect their learning positively.

**Table 8: use of social networking applications (e.g. Facebook, twitter) in learning**

| Scale | SA | A | UD | DA | SDA | Mean |
|---|---|---|---|---|---|---|
| Freq. | 120 | 136 | 24 | 18 | 02 | 4.18 |
| %age | 40.00% | 45.33% | 8.00% | 6.00% | 0.67% | |

Table 9 shows that 66.00% of the students agreed that they used social book-marking tools (e.g. Delicious, Digg), 13.34% students disagreed and 20.67% students remained undecided. The computed mean score was 3.69.

**Table 9: use of social book-marking tools (e.g. delicious, digg) by students**

| Scale | SA | A | UD | DA | SDA | Mean |
|---|---|---|---|---|---|---|
| Freq. | 56 | 142 | 62 | 32 | 08 | 3.69 |
| %age | 18.67% | 47.33% | 20.67% | 10.67% | 2.67% | |

### 7.2. Testing of Hypothesis:

$H_0$: There is no significant difference in the perceptions of male and female students about the use of BL at public sector universities.

$H_1$: There is a significant difference in the perceptions of male and female students about the use of BL for learning purpose.

- Analysis of problem
- $H_0: \mu_1 = \mu_2$
- $H_1: \mu_1 \neq \mu_2$
- $\alpha = 0.05$
- Test statistics: t- test

Referring to table 10, it can be seen that the tabulated value of t = 1.965 with df=298 at $\alpha$=0.05 is less than the computed value of t= 18.443. Therefore, the null hypothesis is rejected, and it is concluded that there is a significant difference between the views of male and female students regarding the implementation of BL at public sector universities in Pakistan.

**Table 10: Students' Responses (t-TEST)**

| Male students ($X_1$) | | Female students ($X_2$) | |
|---|---|---|---|
| $X_1$ | $(X_1)^2$ | $X_2$ | $(X_2)^2$ |
| 52 | 2704 | 62 | 3844 |
| 51 | 2601 | 58 | 3364 |
| - | - | - | - |
| 32 | 1024 | 48 | 2304 |
| 56 | 3136 | 50 | 2500 |

- $\Sigma X_1 = 7554$           $\Sigma X_2 = 7333$
- $N_1 = 152$                    $N_2 = 148$
- $(\Sigma X_1)^2 = 383{,}866$   $(\Sigma X_2)^2 = 384{,}053$

## 8. Discussion and Conclusion

Blended learning gives the opportunity for university students to use new educational technologies. Technology in education was one of the perceived outcomes that students appreciate most in BL, as it is the best way to utilise traditional face to face learning methods. The study concludes that all respondents had a positive inclination towards the use of technology in learning, and almost all respondents agreed that it is helpful, useful and advantageous. Regarding the level of implementation of BL, it was found that there are barriers, such as lack of motivation, technological resources and financial burdens hampering students' use of technology. Students use technology-equipped classrooms and laboratories and have proper assistance provided when any technical problem occurs. Students agreed that the use of technology affects their achievement positively, and it encourages peer collaboration. Teachers encourage and motivate students to use personal devices (e.g. cell phone, mp3 player, PDA) and social networking applications to help and boost their learning.

Technology-equipped classrooms were mostly available in natural science faculties and less in social science faculties. However, students from remote areas face a lot of difficulties in the use of technological sources, and the majority of the student population at general public sector universities belong to remote areas.

## 9. Recommendations

- Universities should develop comprehensive institutional and organisational mechanisms to implement BL for effective teaching of students.
- The university administration should provide extra computing infrastructure (e.g. servers, bandwidth, storage capacity) to faculties and departments so that courses in a blended format can be introduced to the students.
- In the strategic plan for the universities, the BL courses should be well defined and highlighted.
- In policies and planning, the universities' statuaries bodies (i.e. academic council, syndicate, senate) should focus on the implementation of BL in courses.
- The heads of different departments in the universities may be provided with necessary guidelines to implement the already developed organisational mechanisms for the BL mode of learning according to the vision and mission of the university.
- The technology based centralised resource centre should be established to provide technical support and guidance to the students and teachers.
- The learning management system (LMS) should be introduced at department level through the technology based centralised resource centre.
- Conferences and seminars on BL should be organised in collaboration with technologically sufficient institutions e.g. Virtual University Pakistan.
- The separate budgetary heads should be maintained for the purchase and provision of equipment and software needed for BL.
- Keeping COVID-19 in mind, universities should take initiatives that may be appropriate, instead of face to face classroom interaction, when social distance is required.

## Biographies

**Saira Soomro**

*Department of Distance Continuing and Computer Education, Faculty of Education, University of Sindh, Hyderabad, Pakistan, 00923008370324, saira@usindh.edu.pk*

Ms Soomro holds a BS in Computer Science, a master's degree in education and an MPhil in Education from the University of Sindh. She has been an assistant professor in the Department of Distance Continuing and Computer Education, Faculty of Education, University of Sindh, Pakistan, since 2007. She served as a university coordinator for the Pre-STEP and TEP projects funded by international agency USAID in collaboration with the Higher Education Commission, Pakistan. Her areas of interest include e-learning, blended learning, instructional technology and computer-supported collaborative learning.

**Arjumand Bano Soomro**

*Department of Software Engineering, Faculty of Engineering and Technology, University of Sindh, Jamshoro, Sindh, Pakistan, 00923350324434, arjumand@usindh.edu.pk*

Dr Soomro has a PhD in Information Technology from International Islamic University, Malaysia. She holds a master's in education and an MPhil in Information Technology from the University of Sindh. She has a 4-year (Hons) degree in computer and information technology, University of Sindh, Jamshoro, Pakistan. Currently, she is a lecturer for the software engineering department, Faculty of Engineering and Technology, University of Sindh, Jamshoro. Her research areas are ICT in education, blended learning, information systems, software engineering, software teams, team climate and systematic review.
ORCID: 0000-0001-5230-7459

## Tarique Bhatti

*Department of Psychological Testing Guidance & Research, Faculty of Education, University of Sindh, Hyderabad, Pakistan, 00923063031551, tarique.bhatti@usindh.edu.pk*

Dr Bhatti has been associated with the University of Sindh, Jamshoro, since 2009 and is an assistant professor in the Faculty of Education. He completed his PhD in the field of education from the University of Sindh in 2015. He has worked on research projects funded by the international agencies USAID and CIDA. He participated in different professional development activities and developed courses in B. Ed (Hons) 4-year program. His areas of interest include research, psychological testing, guidance and counselling, and teacher training.

## Yonis Gulzar

*Department of Management Information Systems, College of Business, King Faisal University, Al Ahsa, Saudi Arabia, 00966545719118, ygulzar@kfu.edu.sa*

Dr Gulzar is currently an assistant professor at King Faisal University (KFU), Saudi Arabia. He obtained his PhD in Computer Science from the International Islamic University Malaysia in 2018. He completed his master's degree in computer science from Bangalore University, India in 2013. His research interests include database systems, query processing, preference queries, skyline queries, probabilistic and uncertain databases, incomplete data, data integration, location-based social networks (LBSN), recommendation systems and data management in cloud computing. ORCID: 0000-0002-6515-1569